\documentclass[pdftex,twocolumn,epjc3]{svjour3}
\usepackage{amsmath}
\usepackage{amssymb}

\RequirePackage[T1]{fontenc}
\smartqed
\RequirePackage{graphicx}
\RequirePackage{mathptmx}
\RequirePackage{flushend}
\RequirePackage[numbers,sort&compress]{natbib}
\RequirePackage[colorlinks,citecolor=blue,urlcolor=blue,linkcolor=blue]{hyperref}

\journalname{Eur. Phys. J. C}

\begin{document}

\title{Scattering and bound states for the Hulth\'{e}n potential in a
  cosmic string background}

\author{
  Mansoureh Hosseinpour\thanksref{e1,addr1}
  \and
  Fabiano M. Andrade\thanksref{e2,addr2}
  \and
  Edilberto O. Silva\thanksref{e3,addr3}
  \and
  Hassan Hassanabadi\thanksref{e4,addr1}
}

\thankstext{e1}{e-mail: hosseinpour.mansoureh@gmail.com}
\thankstext{e2}{e-mail: fmandrade@uepg.br}
\thankstext{e3}{e-mail: edilbertoo@gmail.com}
\thankstext{e4}{e-mail: h.hasanabadi@shahroodut.ac.ir}

\institute{
  Physics Department,
  Shahrood University of Technology,
  P. O. Box: 3619995161-316,
  Shahrood, Iran\label{addr1}
  \and
  Departamento de Matem\'{a}tica e Estat\'{i}stica,
  Universidade Estadual de Ponta Grossa,
  84030-900 Ponta Grossa, PR, Brazil\label{addr2}
  \and
  Departamento de F\'{i}sica,
  Universidade Federal do Maranh\~{a}o,
  65085-580 S\~{a}o Lu\'{i}s, MA, Brazil\label{addr3}
}

\date{Received: date / Accepted: date}

\maketitle

\begin{abstract}
In this work we study the Dirac equation with vector and scalar
potentials in the spacetime generated by a cosmic string.
Using an approximation for the centrifugal term, a solution
for the radial differential equation is obtained.
We consider the scattering states under the Hulth\'{e}n potential and
obtain the phase shifts.
From the poles of the scattering $S$-matrix the states energies are
determined as well.
\end{abstract}

\section{Introduction}
\label{sec:introduction}

To study the relativistic quantum dynamics of particles with spin under
the influence of electromagnetic fields in curved spacetime we must
consider the modified covariant form of the Dirac equation ($\hbar=c=1$)
\cite{Book.2012.Lawrie,Book.1984.Birrell}
\begin{equation}
  \left\{i\gamma^{\mu}(x)
    \left[ \partial_{\mu}+\Gamma_{\mu}(x)+ i e A_{\mu}(x)
    \right] -M
  \right\} \psi(x) = 0,
  \label{eq:diraccurved}
\end{equation}
where $A_{\mu}$ denotes the vector potential associated with the
electromagnetic field, $\Gamma_{\mu}(x)$ is the spinor
affine connection and $\gamma^{\mu}(x)$ are the Dirac
matrices in the curved spacetime.
The $\gamma^{\mu}\left( x\right)$
matrices are constructed from the standard Dirac matrices in Minkowski
spacetime, which are written in terms of local coordinates, and write
them in terms of global coordinates using the inverse vierbeins
$e_{a}^{\mu}\left( x\right)$ through the relation
\begin{equation}
\gamma^{\mu}(x)=e_{a}^{\mu}(x)\gamma^{a}, \quad (\mu ,a=0,1,2,3),  \label{eq:diracmathe}
\end{equation}
with $\gamma^{a}=\left( \gamma^{0},\gamma^{i}\right)$ being the standard
gamma matrices.
The $\gamma^{\mu}(x)$ matrices also define
the covariant Clifford algebra,
\begin{equation}
\left\{\gamma^{\mu}\left( x\right),\gamma^{\nu}(x)\right\}
=2g^{\mu \nu}\left( x\right).
\end{equation}
If we want the particle to interact with scalar potentials it is more
convenient to write the Dirac equation \eqref{eq:diraccurved} in the
following form
\cite{JHEP.2004.2004.16}:
\begin{equation}
\left\{\alpha^{i}\left[ p_{i}-i\Gamma_{i}-eA_{i}\right] +eA_{0}-i\Gamma
_{0}+\beta M\right\} \psi (x)=E\psi(x),
\label{eq:diraccurved2}
\end{equation}
for $i=1,2,3$, and then introduce the vector and scalar couplings,
\begin{align}
E &\rightarrow E-V\left( r\right),  \label{eq:cb} \\
M &\rightarrow M+S\left( r\right),  \label{eq:cc}
\end{align}
respectively.

It is important to note that these couplings differ
in the manner how they are inserted into the Dirac equation.
In Ref. \cite{Book.2000.Greiner} it was shown how the vector coupling in
\eqref{eq:cb} acts on electron and positron states.
As a result, the potential couples to the charge and a great number of
physical phenomena can be studied through the Dirac equation
\eqref{eq:diraccurved}.
For instance, it is used to study the relativistic quantum motion
of charged spin-0 and spin-1/2 particles in the presence of a uniform
magnetic field and scalar potentials \cite{EPJC.2012.72.2051}, to study the
influence of topological defects on the spin current as well as the spin
Hall effect \cite{PRA.2013.87.032107} and to investigate the role of the
cosmic string on spin current and Hall electric field
\cite{PRD.2014.90.125014}.
On the other hand, contrary to the coupling \eqref{eq:cb}, the scalar
coupling \eqref{eq:cc} acts equally on particles and antiparticles.
Namely, we can add it directly to the mass term of the Dirac equation.
The most common interpretation known in the literature for this coupling
is that of a position-dependent effective mass \cite{JPA.43.325302.2010}. 
This coupling has been used, for example, to study the problem of a
relativistic particle with position-dependent mass in the presence of a
Coulomb and a scalar potential in the background spacetime generated by
a cosmic string \cite{MPLA.2013.28.1350137}. 

These couplings are also used to study various physical models by
including interactions and then to investigate their possible physical
implications on system dynamics.
For example, the Aharonov--Bohm effect
\cite{AoP.1996.251.45,EPJC.2013.73.1}, the Dirac particle in a Morse 
potential \cite{PLB.2011.703.379,PRA.2013.87.052122}, the Dirac equation
for an attractive Coulomb potential in supersymmetric quantum mechanics
\cite{JPA.1985.18.L697}, the Dirac equation with scalar and vector
potentials under the exact spin and pseudospin symmetries limits
\cite{PS.2008.77.045007,JPA.2007.40.263,PS.2007.75.170,PLA.2006.351.379} 
and particles interacting with the Hulth\'en \cite{PLA.2005.344.117} and
Rosen--Morse \cite{JMP.1992.33.643} potentials. 

Among the most important problems that we can study using the couplings
\eqref{eq:cb} and \eqref{eq:cc}, we refer to the scattering problems.
The scattering problems of quantum systems, i.e., the prediction of
reaction probabilities when two objects collide, provide us with a
reliable understanding of the physical system under investigation and
they belong the most effective ways to study the structure of matter.
Great parts of our current theories are built on the basis of scattering 
experiments. 
The best-known example of this claim is the atomic nucleus.
Scattering problems are among the most technically demanding problems in
quantum physics. 
The underlying difficulty lies in the unbounded nature of the wave
function in these problems.
The process of scattering of particles by potentials changes their wave
function by introducing phase shifts.
The study of these phase shifts allow us to predict experimental
observations from the fundamental interactions postulated by the theory.
These studies have already been done in nonrelativistic and relativistic
quantum mechanics
\cite{Book.2000.Greiner,PR.77.699.1950,PRL.44.1559.1980,
PRD.22.1922.1980,IJMPA.30.1550124.2015,PRD.40.1346.1989,
Inproceedings.2004.Cotaescu}.
Nevertheless, the dominant part of these scattering studies is limited
to flat spacetime and the cases regarding the curved spacetime are
actually less frequent.
On the other hand, the dynamics of quantum mechanics on curved spaces in
the presence of topological defects has attracted much attention in
recent years
\cite{EPL.62.306.2003,PLA.188.394.1994,PLA.195.90.1994,
EPL.45.279.1999,JPA.34.5945.2001,EPJC.72.2051.2012,
PRD.91.064034.2015,PRD.93.043545.2016,PRD.95.045005.2017}.

In the present work, we are going to consider the scattering problem of the
Dirac equation produced by the Huth\'{e}n potential in a cosmic string
background.
As will be showed, due to the presence of the centrifugal term in the
radial differential equation, it cannot be solved in an exact manner.
An approximation is used in order to obtain an approximated
solution for the problem.
The cosmic string is a linear defect that changes the topology of the
medium.
This field has been an appealing research field in the past years as not
much is known in comparison with the ordinary Dirac equation in flat
space \cite{PRD.89.27702.2014,PLA.361.13.2007,NPB.328.140.1989,
PRD.85.041701R.2012,AoP.339.510.2013,EPJC.74.3187.2014}.
From the field theory point of view, the cosmic string can be viewed as
a consequence of symmetry breaking phase transition in the early
Universe \cite{Book.2000.Vilenkin}.
Until now some problems have been investigated in
curved spacetime including the one-electron atom problem
\cite{NC.115B.11.2000,Book.1981.Landau,PRD.66.105011.2002}.
The dynamics of non-relativistic particles in curved spacetime is
considered in
\cite{PRL.80.2257.1998,JPA.32.571.1999,ZPB.29.101.1978,
PLA.289.160.2001,EPL.52.1.2000,EPL.45.279.1999,PLA.195.90.1994}
as well.

This paper is organized as follows.
In Sect. \ref{sec:dirac_eq}, we study the covariant Dirac equation in
the spacetimes generated by a cosmic string in the presence of vector and
scalar potentials of electromagnetic field.
We then find special cases of the equation for equal and opposite scalar
and vector potentials.
In Sect. \ref{sec:phase_shift}, we consider the Dirac equation with
the Hulth\'{e}n potential in the context of spin and pseudo-spin
symmetries and obtain the scattering solutions as well as the phase
shifts. 
In Sect. \ref{sec:bound_states}, we derive the scattering $S$-matrix
and from its poles we determine the bound state energies. 
Finally, in Sect. \ref{sec:conclusion}, we present our conclusions.

\section{Dirac equation in cosmic string background}

\label{sec:dirac_eq}

The line element corresponding to the cosmic string spacetime
\cite{CQG.12.471.1995,PRD.47.4273.1993} is
given, in spherical coordinates, by
\cite{PR.121.263.1985,GRG.17.1109.1985} 
\begin{equation}
  ds^{2}=-dt^{2}+r^{2}d\theta^{2}+
  \alpha^{2}r^{2}\sin^{2}\theta d\varphi^{2},
  \label{eq:metric}
\end{equation}
where $t\in (-\infty,\infty)$, $r\in [0,\infty)$, $\theta\in [0,\pi/2]$ and
$\phi\in [0,2\pi]$.
The $\alpha$ parameter in Eq. \ref{eq:metric} is related to the linear
mass $\bar{\mu}$ of the string by $\alpha = 1 - 4 \bar{\mu}$ and it is
defined in the range $(0,1]$.

Now, in order to write the Dirac equation \eqref{eq:diraccurved2} in the
cosmic string spacetime, we must rewrite the Dirac matrices in terms of
global coordinates. 
Additionally, we have to calculate the affine spinorial connection
$(\Gamma_0,\Gamma_i)$. 
The details of this calculation can be found in
\cite{IJMPA.30.1550124.2015,JHEP.2004.2004.16}.
By using the wave function decomposition in the form
\begin{equation}
\psi(x)=e^{-iEt}\chi \left( r,\theta ,\varphi \right) ,
\end{equation}
the Dirac equation in \eqref{eq:diraccurved2} can be written as
\begin{multline}
  \bigg\{
  i\alpha^{r}\partial_{r}+\frac{i\alpha^{\theta}}{r}\partial_{\theta}
  +\frac{i}{\alpha r\sin \theta}\alpha^{\varphi}\partial_{\varphi}\\
  +\frac{i}{2r}\left(1-\frac{1}{\alpha}\right)
  \left(\alpha^{r}+\cot \theta\alpha^{\theta}\right)\\
  -  \gamma^{0}\left[ M+S(r) \right]
  +E-V(r) \bigg\}\chi(r,\theta,\varphi)  = 0,
  \label{eq:diraccp}
\end{multline}
where we have included a scalar $M\rightarrow M+S(r) $ and a
vector $E\rightarrow E-V(r)$ coupling.
Now, assuming that the solutions of Eq. \eqref{eq:diraccp} are of the form
\cite{PRD.66.105011.2002} 
\begin{equation}
  \chi (r,\theta ,\varphi) =
  r^{-{(\alpha-1)}/{2\alpha}}
  \left(\sin \theta \right)^{-{(\alpha-1)}/{2\alpha}}
  F(r) \Theta(\theta) \Phi(\varphi) ,
\end{equation}
we find the radial equation
\begin{multline}
  \bigg\{
    \tilde{\alpha}^{r}p_{r}
    +\frac{i}{r}\tilde{\alpha}^{r}\sigma^{z}
    \boldsymbol{\kappa}_{(\alpha)}+V(r) +
    \left[ M+S(r)\right] \sigma^{z}\bigg\}
    F_{n,\kappa_{(\alpha)}}(r)\\
    = E  F_{n,\kappa_{(\alpha)}}(r),
    \label{eq:diraccnn}
\end{multline}
with \cite{PRD.66.105011.2002}
\begin{equation}
  \tilde{\alpha}^{r}=\left(
    \begin{array}{cc}
      0 & -i \\
      i & 0
    \end{array}
  \right) ,\qquad
  \sigma^{z}=\left(
    \begin{array}{cc}
      1 & 0 \\
      0 & -1
    \end{array}
  \right).
\end{equation}
In Eq. \eqref{eq:diraccnn}, $\boldsymbol{\kappa_{(\alpha)}}$ represents the
generalized spin--orbit operator in the spacetime of a cosmic string whose
eigenvalues are given by
\begin{equation}
\kappa_{(\alpha)}=\pm \left[ j_{(\alpha)}+\frac{1}{2}\right] =\pm \left[
j+m\left(\frac{1}{\alpha}-1\right) +\frac{1}{2}\right],
\label{eq:generalspin}
\end{equation}
with $j_{(\alpha)}$ representing the eigenvalues of the generalized total
angular momentum operator.
The operator $\boldsymbol{\kappa}_{(\alpha)}$
is given by
\begin{equation}
  \sigma^{z}\boldsymbol{\kappa}_{(\alpha)}=
  \boldsymbol{\tilde{\alpha}}\cdot\mathbf{L}_{(\alpha)}+1,
\end{equation}
where $\mathbf{L}_{(\alpha)}$ is the generalized angular momentum operator
in the spacetime of the cosmic string,
\begin{equation}
L_{(\alpha)}^{2}Y_{\ell_{(\alpha)}}^{m_{(\alpha)}}(\theta ,\varphi
)=\ell_{(\alpha)}\left(\ell_{(\alpha)}+1\right) Y_{\ell_{(\alpha
)}}^{m_{(\alpha)}}(\theta ,\varphi),
\end{equation}
with $Y_{\ell_{(\alpha)}}^{m_{(\alpha)}}(\theta ,\varphi)$ being the
generalized spherical harmonics and $m_{(\alpha)}$ and $\ell_{(\alpha)}$
not necessarily being integers.
In particular $m_{(\alpha)}=m/\alpha $, where
$m=0,\pm 1,\pm 2,\ldots $, $\alpha \in (0,1]$ and $\ell_{(\alpha
)}=n+m_{(\alpha)}=\ell +m(1/\alpha -1)$, $\ell =0,1,2,\ldots ,n-1$.
Here $
\ell $ and $m$ are, respectively, the orbital angular momentum quantum
number and the magnetic quantum number in the flat space (i.e., for
$\alpha=1$), and $n$ is the principal quantum number.

By choosing the radial wave function as \cite{PRD.66.105011.2002}
\begin{equation}
F(r) =\frac{1}{r} \left(
\begin{array}{c}
-if(r) \\
g(r)
\end{array}
\right) ,
\end{equation}
we obtain the coupled equations
\begin{align}
  -i\left[ E-M-\left(S+V\right) \right] f(r)
  +\frac{dg(r)}{dr}
  +\frac{\kappa_{(\alpha)}}{r}g(r) = {} & 0,\\
  -i\left[ E+M+S-V\right] g(r)
  +\frac{df(r)}{dr}
  +\frac{\kappa_{(\alpha)}}{r}f(r) = {} & 0.
\end{align}
Let us now consider the special case of $S(r)=V(r)$ (exact spin symmetry
limit) and $S(r) =-V(r)$ (exact pseudo-spin symmetry limit).
After
elimination of one component in favor of the other, for $S(r)=V(r)$, we have
\begin{multline}
  -\frac{d^{2}f(r)}{dr^{2}}
  +\bigg[
  \frac{\kappa_{(\alpha)}\left(\kappa_{(\alpha)}-1\right)}{r^{2}} \\
  - \left(E-M-2V(r)\right) \left(E+M\right)
  \bigg] f(r) = 0.
  \label{eq:sspin}
\end{multline}
Additionally, for the case $S(r) =- V(r)$, we obtain
\begin{multline}
  -\frac{d^{2}g(r)}{dr^{2}}
  + \bigg[
  \frac{\kappa_{(\alpha)}\left(\kappa_{(\alpha)} +1\right)}{r^{2}} \\
  -\left(E+M-2V(r)\right) \left(E-M\right)
  \bigg] g(r) =0.
 \label{eq:spspin}
\end{multline}
Thus comparing Eqs. \eqref{eq:sspin} and \eqref{eq:spspin}, we can see that the
solution for the case $S(r)=-V(r)$, can be obtained from the solution for
the case $S(r)=V(r)$ with the replacements $\kappa_{(\alpha)}-1 \to
\kappa_{(\alpha)}+1$ and $M \to -M$.
Therefore we shall only deal with the latter because
the results for the former can be obtained in a straightforward manner
by using the above replacements.

\section{Scattering analysis}
\label{sec:phase_shift}

In this work, we are interested in considering the Hulth\'{e}n
potential, which has remarkable applicabilities because of its
short-range nature. 
It should be noted that this potential is a special case of the
Eckert potential \cite{JPA.40.10535.2007}.
Therefore, we are going to investigate scattering state solutions of
the Dirac equation in the presence of the Hulth\'{e}n potential,
\begin{equation}
V(r) =-\xi\frac{\omega}{e^{\omega r}-1},  \label{eq:hulthen}
\end{equation}
where $\omega$ is the screening parameter and $\xi$ is a positive
constant.
When $V(r)$ is used to describe atomic phenomena, $\xi$ is
interpreted as $Ze^{2}$, with $Z$ the atomic number.
In this step, we want to evaluate phase shifts and normalization factor
for the pseudo-spin symmetry limit (i.e., $S(r)=V(r))$.
Thus inserting Eq. \eqref{eq:hulthen} into \eqref{eq:sspin}, we obtain
\begin{multline}
  \bigg[
  -\frac{d^{2}}{dr^{2}}
  +\frac{\kappa_{(\alpha)}(\kappa_{(\alpha)}-1)}{r^{2}}
  -2 (E+M) \xi\frac{\omega e^{-\omega r}}{1-e^{-\omega r}}
  \bigg]f(r)\\
  = k^{2}f(r),
  \label{eq:ndiff}
\end{multline}
where $k^{2}=E^{2}-M^{2}$.
It is worthwhile to note here that, for small values of $\omega$, the
Hulth\'{e}n potential behaves like the Coulomb potential, consequently
Eq. \eqref{eq:ndiff} turns into the Bessel equation
\cite{CMP.124.229.1989,NPB.349.207.1991,AoP.202.271.1990,NPB.353.237.1991}.
In contrast with the Coulomb potential, unfortunately, the Dirac
equation for the Hulth\'{e}n potential cannot be solved analytically due
to the presence of the centrifugal term.
In this manner, it is necessary to use some approximation.
Considering small values of $\omega$, a common approximation for the
centrifugal term is
\begin{equation}  \label{eq:approx}
  \frac{1}{r^{2}} \approx
  \frac{\omega^{2}e^{-\omega r}}{(1-e^{-\omega r})^{2}}.
\end{equation}
The above approximation, as we shall see, leads us to a solvable
differential equation  \cite{PLA.368.13.2007,JPA.40.10535.2007}.
Therefore, by using the approximation in Eq. \eqref{eq:approx},
followed by the change of variable $y=1-e^{-\omega r}$, we can write the
Eq. \eqref{eq:ndiff} in the form
\begin{multline}
  \bigg[
    \frac{d^{2}}{dy^{2}}-\frac{1}{1-y}\frac{d}{dy}
    -\frac{\kappa_{(\alpha)}(\kappa_{(\alpha)}-1)}{y^{2}(1-y)}
    +\frac{\eta^{2}}{y(1-y)}\\
    +\frac{\varepsilon^2}{(1-y)^{2}}
    \bigg] f(y) = 0,
  \label{eq:newhu}
\end{multline}
in which
$\eta^{2}={2(E+M)\xi}/{\omega}$ and $\varepsilon^{2}=k^{2}/\omega^{2}$.
Equation \eqref{eq:newhu} can be turned into a well-known differential
equation if we choose
\begin{equation}
f(y) =y^{\gamma}\left(1-y\right)^{\nu}h(y),  \label{eq:newf}
\end{equation}
as the solutions, where $\nu$ and $\gamma$ are arbitrary constants to be
determined.
Therefore, substituting Eq. \eqref{eq:newf} into Eq.
\eqref{eq:newhu} we obtain
\begin{multline}
\bigg[
  \frac{d^2}{dy^2}
  +\frac{2\gamma - (1 + 2\nu + 2\gamma)y}{y(1-y)}\frac{d}{dy}
  +\frac{\nu^2+\varepsilon^2}{(1-y)^2}
  \\
  +\frac{\eta^2-2\gamma\nu-\gamma^2}{y(1-y)}
   +\frac
  {\gamma(\gamma-1)-\kappa_{(\alpha)}(\kappa_{(\alpha)}-1)}
  {y(1-y)^2}
  \bigg]
  h(y)  = 0.
 \label{eq:newode}
\end{multline}
We determine the parameters $\nu$ and $\gamma$ by imposing that the
coefficients of the terms  $1/(1-y)^{2}$ and $1/[y(1-y)^{2}]$ vanish
identically.
In this manner, we have
\begin{align}
  \nu = \pm \frac{i k}{\omega},\qquad
  \gamma = \kappa_{(\alpha)} \text{ or }
  \gamma = 1-\kappa_{(\alpha)}.
\end{align}
This set of parameters leads us to the same solution for
Eq. \eqref{eq:newode} and we are free to choose one set.
Thus choosing $\nu=i k/\omega$ and $\gamma=\kappa_{(\alpha)}$ this leads
to
\begin{equation}
\left\{y(1-y) \frac{d^{2}}{dy^{2}} +[\eta_{3}-(1+\eta_{1}+\eta_{2} y]\frac{
d}{dy} -\eta_{1}\eta_{2} \right\} h(y) = 0,  \label{eq:finalode}
\end{equation}
where
\begin{align}
\eta_{1} = {} & \gamma + \nu + \sqrt{\eta^{2}+\nu^{2}}= \kappa_{(\alpha)} +
\frac{i k}{\omega} +\sqrt{\frac{2(E+M)\xi}{\omega}-\frac{k^{2}}{\omega^{2}}},
\notag \\
\eta_{2} = {} & \gamma + \nu - \sqrt{\eta^{2}+\nu^{2}}= \kappa_{(\alpha)} +
\frac{i k}{\omega} -\sqrt{\frac{2(E+M)\xi}{\omega}-\frac{k^{2}}{\omega^{2}}},
\notag \\
\eta_{3} = {} & 2\gamma = 2\kappa_{(\alpha)}.
\end{align}
Equation \eqref{eq:finalode}, has the form of a hypergeometric
differential equation \cite{Book.1981.Landau},
\begin{equation}
h(y) ={}_{2}F_{1}(\eta_{1},\eta_{2},\eta_{3};y).
 \label{eq:solhu}
\end{equation}
Therefore, the radial wave functions can be written as
\begin{equation}
f(r) = N (1-e^{-\omega r})^{\ell} e^{ikr}
{}_{2}F_{1}\left(\eta_{1},\eta_{2},\eta_{3};1-e^{-\omega r}\right).
\label{eq:sgb}
\end{equation}

Now, in order to obtain the scattering phase shift and the normalization
factor we write the asymptotic form of the above  wave function.
For this purpose we  use the properties of the
hypergeometric functions \cite{Book.2007.Gradshteyn} and the asymptotic behavior of \eqref{eq:sgb} to
$r\rightarrow \infty $ \cite{IJMPA.30.1550124.2015}
\begin{align}
  f(r) \sim {}
  & 2 N
    \left[ \Gamma \left(\eta_{3}\right) \right]^{2}
    \left|
    \frac
    {\Gamma(\eta_{3}-\eta_{1}-\eta_{2})}
    {\Gamma(\eta_{3}-\eta_{1})\Gamma(\eta_{3}-\eta_{2})}
    \right|\\
  &                
    \times \sin \left(k r+\frac{\pi}{2}+\delta\right),
\label{eq:solgn}
\end{align}
where $N$ is a normalization constant.
Recalling the boundary condition for $r\rightarrow \infty$ imposed in Ref.
\cite{Book.1981.Landau} as
\begin{equation}
f (r) \sim 2 \sin \left(kr-\frac{\ell \pi}{2}+\delta_{\ell}\right) ,
\end{equation}
and comparing with Eq. \eqref{eq:solgn}, the phase shift and the
normalization factor can be found.
The result is
\begin{equation}
  \delta_{\ell}=
  \frac{\pi}{2} (\ell+1)+
  \arg
  \left[
    \frac
    {\Gamma(\eta_{3}-\eta_{1}-\eta_{2})}
    {\Gamma(\eta_{3}-\eta_{1})\Gamma(\eta_{3}-\eta_{2})}
  \right],  \label{eq:shift}
\end{equation}
and
\begin{equation}
  N =
  \frac{1}{\left[\Gamma \left(\eta_{3}\right) \right]^{2}}
  \left\vert
    \frac
    {\Gamma \left(\eta_{3}-\eta_{1}-\eta_{2}\right)}
    {\Gamma \left(\eta_{3}-\eta_{1}\right)
      \Gamma \left(\eta_{3}-\eta_{2}\right)}
  \right\vert .
 \label{eq:cn}
\end{equation}
It can be seen that when $\alpha = 1$ the results are the same as the
Dirac equation in Minkowski spacetime.
In this case, the generalized spin--orbit
operator in Eq. \eqref{eq:generalspin} reduces to
\begin{equation}
\kappa =\pm \left(j+\frac{1}{2}\right),
\end{equation}
where $\gamma = \kappa$.
Therefore we can obtain the phase shifts and the normalization factor in
flat spacetime in this case if we rewrite the Dirac equation in the flat
spacetime by standard Dirac matrices. 
As expected our result is the limit of Eqs. \eqref{eq:shift} and
\eqref{eq:cn} when $\alpha = 1$.

\section{Bound states analysis}
\label{sec:bound_states}

The Hulth\'{e}n potential also admits bound state solutions.
In order to
find the bound states energies, let us analyze the scattering $S$-matrix.
It is well known that poles of the $S$-matrix in the upper half of the complex
plane are associated with the bound states.
Using Eq. \eqref{eq:shift}, the $S$-matrix can be written as
\begin{align}
  S_{\ell} = {}
  &
    e^{2i\delta_{\ell}} \nonumber\\
  = {}
  &
    e^{i\pi(\ell+1)} e^{2i\arg{[\Gamma(\eta_3-\eta_1-\eta_{2})]}}
  \nonumber\\
  &  
    \times
    e^{-2i\arg{[\Gamma(\eta_3-\eta_1)]}}e^{-2i\arg{[\Gamma(\eta_3-\eta_2)]}}.
\end{align}
Therefore, the poles of the $S$-matrix are given by the poles of the gamma
functions $\Gamma(\eta_3-\eta_1)$ and $\Gamma(\eta_3-\eta_2)$.
In this manner, based on the relations $\eta_{3}-\eta_{2}=\eta_{1}^{*}$,
$\eta_{3}-\eta_{1}=\eta_{2}^{*}$ and
$\eta_{3}-\eta_{2}-\eta_{1}=(\eta_{1}+\eta_{2}-\eta_{3})^{*}=2ik/\omega$,
we are looking for the poles of
\begin{equation}
  \Gamma \left(\kappa_{(\alpha)}-\frac{i k}{\omega}
    \pm \sqrt{\frac{2(E+M)\xi}{\omega}-\frac{k^{2}}{\omega^{2}}} \right).
\end{equation}
The gamma function $\Gamma(z)$ has poles at $z=-n$, where $n$ is a
non-negative integer.
Then the bound state energies are given by
\begin{equation}
  k^2 \equiv E^2-M^2 =
  - \frac {\left[(n+\kappa_{(\alpha)})^{2}\omega-2(E+M)\xi\right]^2} {4(n+\kappa_{(\alpha)})^2},
\end{equation}
whit $n=0,1,2,\ldots$.

\section{Conclusion}
\label{sec:conclusion}

In this work we considered the Dirac equation in curved spacetime and
the topology of spacetime in order to describe physics of the system in
the presence of the gravitational fields of a cosmic string.
We obtained the solution of the Dirac equation in the curved spacetime
by considering vector and scalar potentials.
We considered the scattering states of the Dirac equation under the
Hulth\'{e}n potential and obtained as scattering phase shifts.
From the poles of the scattering $S$-matrix we determined the bound
state energies.
When $\alpha=1$, we recover the general solution of the Dirac equation
in usual spherical coordinates, as we should.

\section*{Acknowledgments}
With great pleasure, the authors thank the kind referee for helpful
comments.
FMA acknowledges funding from the Conselho Nacional de
Desenvolvimento Cient\'{i}fico e Tecnol\'{o}gico (CNPq), Grants No.
460404/2014-8 and No. 311699/2014-6.
EOS acknowledges funding from CNPq, Grants No. 482015/2013-6,
No. 306068/2013-3, No. 476267/2013-7, FAPEMA Grant No. 01852/14 (PRONEM)
and FAPES.

\bibliographystyle{spphys}

\end{document}